# Amplitude-Multiplexed readout of single photon detectors based on superconducting nanowires

*Alessandro Gaggero[1,‡,*], Francesco Martini[1,2,‡], Francesco Mattioli[1], Fabio Chiarello[1], Robert Cernansky[2], Alberto Politi[2] and Roberto Leoni[1]*

[1] Istituto di Fotonica e Nanotecnologie – CNR, Via Cineto Romano 42, 00156 Roma, Italy.

[2] Department of Physics and Astronomy, University of Southampton, Southampton SO17 1BJ, UK.

KEYWORDS NbN, superconducting-nanowire, photonic circuits, single photon, silicon-nitride, NbN, multiplexing.

ABSTRACT The realization of large-scale photonic circuit for quantum optics experiments at telecom wavelengths requires an increasing number of integrated detectors. Superconductive nanowire single photon detectors (SNSPDs) can be easily integrated on chip and they can efficiently detect the light propagating inside waveguides. The thermal budget of cryostats poses a limit on the maximum number of elements that can be integrated on the same chip due to the thermal impact of the readout electronics. In this paper, we propose and implement a novel scheme able for an efficient reading of several SNSPDs with only one readout port, enabling the realization of photonic circuits with a large number of modes.

Single photons are promising candidates as quantum bits (qubits) for quantum information applications due to their low decoherence and ease of transmission, both in free space and by means of optical fibers. Small- and medium-scale quantum computing[1] and simulation[2] can be realized combining single-photon sources, single-photon detectors and linear optics. In the last ten years many efforts have been devoted to develop an integrated platform containing all functionalities needed to achieve the tantalizing "quantum supremacy" Several experiments have been performed or proposed exploiting photonic integrated circuits (PICs), including boson sampling, quantum walk and quantum simulation[3-7]. The increasing complexity of the experiments reflects directly in the PIC architectures, requiring an always-increasing number of integrated components. All of the aforementioned experiments have been performed with hybrid setups, i.e. external sources and detectors, and a clear path to achieve a complete integrated platform is still lacking. SNSPDs are the only detectors[8] that showed an on-chip integration feasibility with outstanding performances in terms of detection efficiency, dark count rate and timing resolution in the infrared wavelength range[9-16]. The increasing PIC complexity requires the integration of several tens of SNSPDs, posing new challenges related to the simultaneous readout of different channels. The use of dedicated readout electronics for each detector channel is not a practical solution and has a tremendous impact on the thermal budget allowed by cryostats. To overcome these limitations, several multiplexing schemes have been proposed based on different approaches. A row-column multiplexing scheme enables the readout of planar 2D array $N^2$ detectors with only 2N channels[17]. Another proposal is based on frequency multiplexing and uses different RF resonators to read out an SNSPD array, where the switching of a portion of the nanowire causes a shift in the resonator operating frequency. Although with this technique it is possible to read out a huge numbers of detectors, each RF tone needs a demultiplexing circuit thus limiting the maximum filling factor achievable by the array[18]. Time domain approaches have been also exploited, where, using a time-tagged multiplexing scheme, the signals coming from two SNSPDs were separated in time using a delay line[19]. This approach requires only a single readout line, but has no photon number resolution and the overall dimensions of the array are dictated by the design of the delay line. Another time tagging scheme was employed to demonstrate a single photon imager using a continuous nanowire delay line. This device was able to resolve the position of absorbed single photons[20]. More recently, another implementation based on time multiplexing was reported[21], where the authors show a two-terminal detector based on a superconducting nanowire microstrip transmission line. The transmission line works as an array that can resolve the position of more than one photon and naturally acts as a coincidence counter. However, to resolve the position of two photons simultaneously, the reference time that generates the double firing event is needed. In addition, a maximum photon number resolution of up to four photon was demonstrated using a complex post processing of the array voltage pulse.

In this work, we implemented a new and simple scheme that is able to read out multiple SNSPDs with only one coaxial cable without the requirement of complex post processing. Our approach consists of a novel version of a spatially multiplexed PNRD (photon number resolving detector)[22-24]. The readout scheme (see Figure 1a) is based on the spatial multiplexing of N active elements consisting of a SNSPD in parallel with an on-chip AuPd resistor, of resistance $R_i$. When a photon is absorbed in an active element a normal resistance $R_n$ appears in the superconductive nanowire and, being $R_n \gg R_i$ by design, all the bias current is diverted to the parallel resistance[25, 26]. The position of the photon-absorption event is then encoded in the voltage amplitude of the pulse. Due to the compact readout scheme, we were able to show a proof of concept based on two-element array integrated in a silicon nitride ($Si_3N_4$) PIC. In addition to retrieve the specific properties of each SNSPD, the circuit is intrinsically able to implement $g^{(2)}(\tau)$ measurements[27, 28]. Combining our approach with the use of a cryogenic amplifier, tens of detectors can be read with a single coaxial cable with a minimum thermal impact on the operating temperature.

In order to show the resolution capability of our approach, we design the parallel resistances $R_1$ and $R_2$ as $R_2 = 2 \cdot R_1$. The detectors $D_1$ and $D_2$ (Figure 1c and d), connected in series, are integrated on the PIC (Figure 1b) and the nanowires length is chosen to be 30 µm, ensuring a good photon absorption up to 77% in the NbN nanowires at wavelength 1550 nm. The thickness of the $Si_3N_4$ layer (350 nm) and the width (1.9 µm) of the waveguides are chosen to minimize the expected optical losses while maintaining the absorption in the nanowires as high as possible. The device fabrication exploits four steps of electron beam lithography (EBL). The first step is used to define the integrated PIC on the $Si_3N_4$ layer that is deposited by PECVD on a 2 µm-thick thermal $SiO_2$[29]. The waveguide pattern (Figure 1 b) is written on a 400 nm-thick CSAR resist and is then transferred to the $Si_3N_4$ using an inductively coupled plasma (ICP) reactive ion etching (RIE). Successively, a 6 nm-thick NbN film is sputtered on the PIC by DC magnetron sputtering at T= 550 °C in a gas mixture of $N_2$+Ar (with 22% $N_2$). Using a second step of EBL, we define the Ti/Au electric contacts on PMMA via lift-off (10/60 nm thick, respectively). The NbN nanowires (80 nm wide) are then patterned on 180 nm thick HSQ resist used as etching mask for the removal of the unwanted NbN (see Figure 1 c and d). Finally, the AuPd resistances are fabricated via lift-off of a Ti (5nm)/AuPd (85 nm) film.

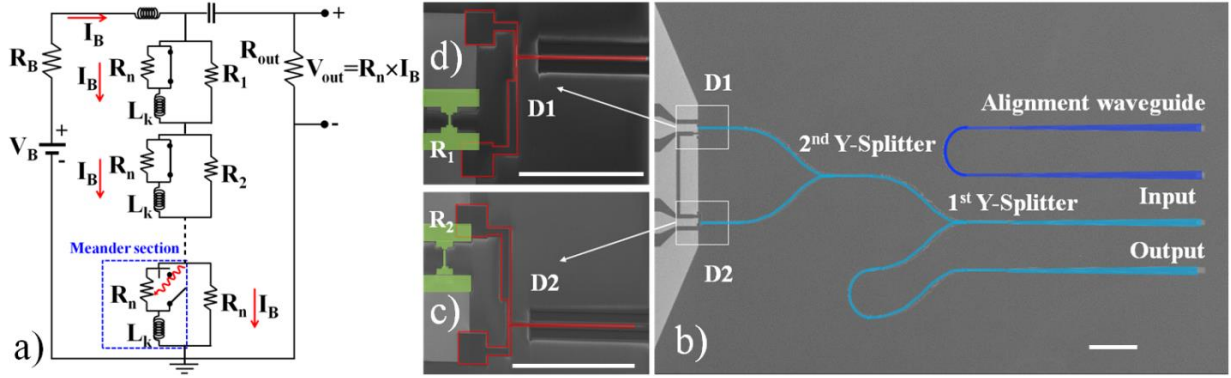

**Figure 1.** a) Sketch of the electrical scheme used to read the pulsed signal at the output of a N-element array. Each element of the array is made by a SNSPD and has a resistance Ri in parallel. A bias-tee is used to bias and read out the array. (b) Scanning electron micrograph of the SNSPD array consisting in two detectors ($D_1$ and $D_2$) integrated on top of a PIC made of two 50:50 Y splitters and input/output ports realized with grating couplers, the scale bar corresponds to 125 µm. Enlarged view of $D_1$ (c) and $D_2$ (d.) In green are highlighted the parallel resistance $R_1$ and $R_2$, respectively, the scale bars are 30 µm.

To avoid the use of expensive and bulky cryogenic positioners, we align and glue a Pyrex fiber array (FA), composed of 6 single mode fibers to the ports of the PIC, obtaining in this way a fast and reliable light coupling. The alignment is obtained at room temperature by maximizing the light at the control port of the photonic circuit (see Figure 1a) by using a 6-axis manipulator for the FA. The coupling efficiency of the grating coupler is about 15% at room temperature. Successively, the chip is mounted on a GM refrigerator operating at a temperature T=2.9 K. During the sample cooling, the transmitted optical power decreases due to the different thermal contractions between the Pyrex (FA) and the silicon substrate, causing a reduction of the coupling efficiency down to ~2%. The readout electronics implements a chain of two bias Tee and two RF amplifiers with a gain of 49 dB in the bandwidth 0.1-500 MHz. The NbN film has a 9 K critical temperature and a critical current density of 3.9 MA/cm$^2$ at 2.9 K. The travelling photons at 1550 nm wavelength, generated by a 10 ps pulsed laser, are injected through the input port of the PIC.

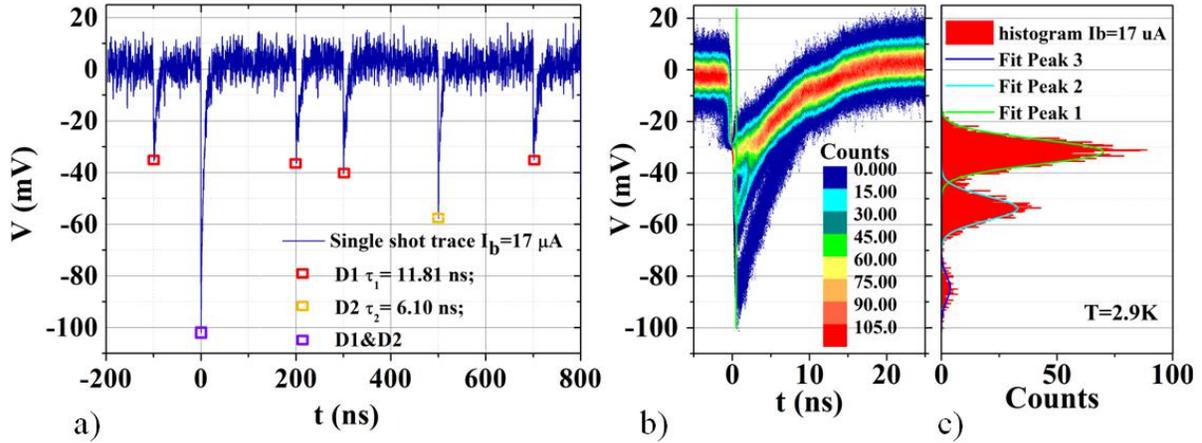

**Figure 2.** a) Single trace of the array output signal showing peaks with three different heights due to the firing of $D_1$ (with $R_1$ in parallel, red squares), $D_2$ (with $R_2$ in parallel, yellow squares) and the simultaneous firing of $D_1$ and $D_2$ (purple squares) at $I_B$=17 μA. The exponential fit of the decay time provides a value of $\tau_1$= 11.81 ns and $\tau_2$= 6.10 ns for the peaks due to $D_1$ and $D_2$, respectively. b) Persistence map of the transient response of the array under the illumination of a 10 ps pulsed laser with 10 MHz repetition rate and an average photon number per pulse <μ> = 1. c) Histogram of pulse counts as function voltage, obtained from the persistence plot using data of the slice taken at 700 ps (green vertical line in b) and its fit with 3 Gaussian curve (continuous lines).

Figure 2 a) shows a single shot trace of the array acquired with a 4 GHz-bandwidth oscilloscope and a laser repetition rate of 10 MHz. Three different pulse amplitudes, related to the value of $R_1$, $R_2$ and $R_1$+ $R_2$, are observed due to the firing events of $D_1$ and $D_2$. The higher pulse is due to the simultaneous firing of $D_1$ and $D_2$, attesting a coincidence event. Using an exponential fit for the pulses, it is possible to evaluate the recovery time of the two peaks related to $R_1$ and $R_2$, corresponding respectively to $\tau_1$= 11.81 ns and $\tau_2$= 6.10 ns, in agreement with a decay time proportional to the inverse of the resistance ($\tau_i$=$L_k$/$R^{30}$, being $L_k$ equal for both the detectors). Figure 2b) displays the persistence map of the array output signal, showing the three amplitude levels, taken with a bias current of 17 μA. For this measurement, we chose to work with an average photon number $<\mu> = 1$ to increase the probability of a coincidence event, that otherwise would be very low in the single photon regime. Using this plot it is possible to reconstruct the histogram of the amplitude probabilities (see Figure 2 c) considering a temporal slice taken at 700 ps (green line of Figure 2b). Using a Gaussian fit we retrieve the position of the three maxima, corresponding to $V_1$= (-31.8± 9.6) mV, $V_2$= (-53.7± 9.4) mV and $V_3$= (-84.6± 9.8) mV. From the peak voltage positions and considering $I_B$= 17 μA, we found $R_1$= 7.81 Ω and $R_2$= 14.98 Ω, in good agreement with the expected values of $R_1$ and $R_2 \approx 2 \cdot R_1$.

The inset of Figure 3a) shows the on-chip detection efficiency (ODE) and dark count rate (DCR) of the array as function of the bias current, measured under the illumination of a pulsed laser with λ= 1550 nm, pulse width of 10 ps and a repetition rate f= 1 MHz. The power of the laser was attenuated down to an average photon per pulse $<\mu> = 0.1$ corresponding to the single photon regime. The array has a maximum ODE of about 24% at a DCR of ~3 kHz and an ODE 11% at 10 Hz of DCR (inset of Figure 3a). The effect of the different pulse amplitudes generated by $D_1$ and $D_2$ is clearly illustrated recording the pulse count rate in function of the voltage trigger level of the frequency counter from -100 mV to 20 mV (see Figure 3 a). We can clearly individuate three levels in the count rate. The first plateau level (in the range between -10 mV to -30 mV) includes the counts due to $D_1$ (pulses with the lowest amplitude), $D_2$ and those due to the coincidences coming from the simultaneous firing of both $D_1$ and $D_2$. The second plateau (in the range between -35 mV to -55 mV) takes into account pulses coming from $D_2$ and from the simultaneous firing of the two detectors, while the third plateau (in the range between -65 mV to -90 mV) is only due to the coincidence events. The plot in Figure 3a) is obtained by subtracting to the counts, the dark count events, measured with the laser off, for each bias current. In the range from -20 mV to 0 mV, the counts due to the firing of $D_1$ and $D_2$ are dominated by the contribution of the electronic noise that is eliminated with the subtraction of the dark counts. Considering that 105 photons/s are coupled at the input of the second 50:50 Y-splitter, from Figure 3a) we can extrapolate the relative ODE of $D_1$ and $D_2$ that correspond to 30.4% and 15.6%, respectively (see Figure 3b). These efficiencies are underestimated as no losses are taken into account due to the last part of the photonic circuit (from the second Y-splitter to the detectors).

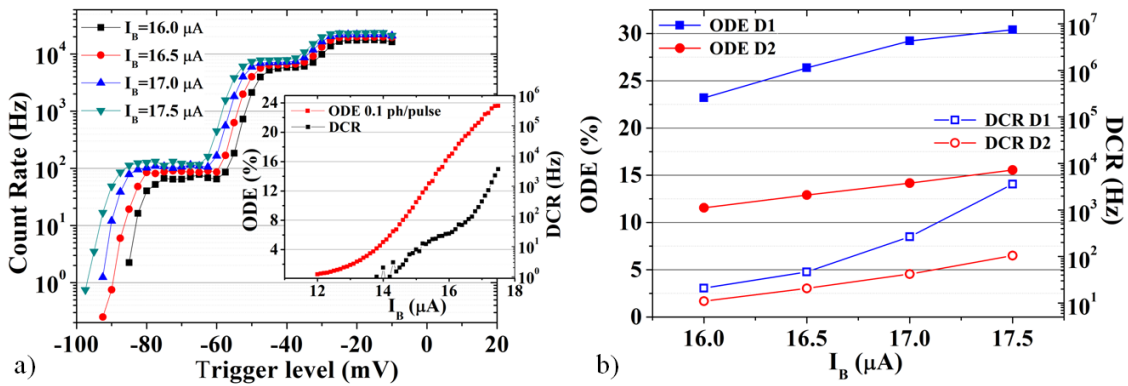

**Figure 3.** a) Pulse count rate of the array vs trigger voltage level of the counter measured at different bias currents for 105 photons/s photon flux coupled to the input of the second Y splitter. Inset: ODE of the array (red squares) and dark counts rate (DCR, black squares) as a function of the bias current, taken at T=2.9 K. b). ODE and DCR of individual $D_1$ and $D_2$, respectively.

The scalability of our approach depends on three main parameters: the input resistance of the readout electronics $R_{out}$ (generally coincides with the RF amplifier impedance), the maximum resistance value $R_M$ allowed by the system and the signal to noise ratio (SNR). In our setup $R_M$ is limited by the 50 $\Omega$ input resistance of the RF amplifier being $R_{out}$ in parallel with the SNSPD array (see Figure 1 a). This poses a limit to the maximum pulse amplitude $V_{max}=I_B \cdot R_M$. In the single photon regime, only one detector is firing at once, implying that the only selection rules that are required for the choice of the set of resistances are $R_i \neq R_{i+1}$ and $\Delta R = R_{i+1} - R_i$ large enough to produce voltage signals greater than the amplifier noise. Therefore, the maximum number of channels readable with a single coax cable is Ch= $R_M/\Delta R$. Being $R_M$= 50 $\Omega$, the scalability of our scheme can be increased by reducing $\Delta R$ as low as possible.

Table 1 summarizes three cases, the first row takes into account all the parameters we measured in our experiment, the second row describes how we can improve our scalability using a commercially available cryogenic RF amplifier[31], while the third one refers to an amplifier with an input impedance of 1 k$\Omega$.

The fit of the histogram (Figure 2 c) provides a variance 2$\sigma$= 7.4 mV for all three Gaussian peaks, giving a noise resistance $\Delta R_{2\sigma}\approx$ 1.6 $\Omega$ (amplifier gain G= 49 dB), in close agreement with $\Delta R_{th}$= 1.2 $\Omega$ calculated from the Johnson-Nyquist voltage noise that is expected for an RF amplifier working at room temperature T= 300 K[32]. As shown in the second row of Tab. 1, a reduction of a factor 15 of the amplifier working temperature corresponds to an increase of SNR of a factor ~ 4. The SNR can be improved also increasing the bias current $I_B$ by improving the film quality, by tuning the geometric parameters of the nanowires or by connecting several nanowires in parallel[33, 34]. In the latter case, we expect an $I_B$ increase of at least a factor two.

| $R_{out}$ ($\Omega$) | T (K) | B (MHz) | $I_B$ ($\mu$A) | $\Delta R_{th}$ ($\Omega$)[32] | $\Delta R_{2\sigma}$ ($\Omega$) | $\Delta R_{6\sigma}$ ($\Omega$) | $R_M$ ($\Omega$) | Ch 6$\sigma$ | 2$\sigma$ |
|---|---|---|---|---|---|---|---|---|---|
| 50 | 300 | 0.1-500 | 17 | 1.20 | 1.60 | 4.70 | 50 | 10 | 30 |
| 50 | 20 | 0.1-500 | 17 | 0.31 | 0.41[35] | 1.20[35] | 50 | 41 | 121 |
| 1000 | 20 | 0.1-500 | 17 | 1.40 | 1.90[35] | 5.70[35] | 200 | 35 | 105 |

**Table 1**: Figures of merit regarding the array readout scalability. For a selected RF amplifier, $R_{out}$ is its input resistance, T the working temperature and B the bandwidth. $\Delta R_{th}$ is the noise resistance due to thermal noise[32] while $\Delta R_{2\sigma}$ and $\Delta R_{6\sigma}$ are the resistance step values considering an experimental noise of 2$\sigma$ and 6$\sigma$ given by the width of the Gaussian fits. The $R_M$ is the maximum resistance allowed by the system and Ch is the number of channels that can be implemented for each electronic configuration in single photon regime in the cases $\Delta R_{2\sigma}$ and $\Delta R_{6\sigma}$.

Selecting an appropriate resolution determined by the step resistances, $\Delta R_{2\sigma}$ and $\Delta R_{6\sigma}$, we can identify the firing element of the array with 68.3% probability and 99.7%, respectively, where the choice strongly depends on the application. For example, if detectors would be used as pixels for spectroscopy or imaging, an uncertainty in the precise position of the firing detector might be tolerated. When error rates have to be minimized, like in quantum computation, it is possible to assign the values of $R_i$ in a strategic way and making use of additional delay lines to separate in time pulses with close voltage amplitudes and thus reject erroneous events.

In the single photon regime and considering that the maximum pulse amplitude is $V_M = I_B \cdot R_M$, where $R_M = 50\ \Omega$, we can calculate the maximum number of elements that can be read, given by $Ch_{6\sigma} = 50/\Delta R_{6\sigma} = 10$ channels at 300 K and $Ch_{6\sigma} = 41$ at 20 K. If we consider a less stringent step resolution of $R_{2\sigma}$, the maximum number of channels rises to $Ch_{2\sigma} = 30$ at 300 K and $Ch_{2\sigma} = 121$ at 20 K.

In the multi photon regime, the firing of different channels occurs simultaneously increasing the number N of Gaussian peaks to be discriminated[36]. For example, the number N of levels required to resolve the position of two elements firing simultaneously is $N = Ch(Ch-1)/2$[36]. In the case of a $50\ \Omega$ RF amplifier at T=300 K (first row of Table 1) the maximum number of channel that can be read in a two-photons regime is $Ch_{2\sigma} \approx 8$ that becomes ~16 by cooling the amplifier at 20 K. However, the lack of linearity due to the parallel with $50\ \Omega$ input impedance of amplifier complicates the selection rules for the resistance steps. The linearity requirements in the multi photons regime can be met by implementing a high-impedance cryogenic readout (row 3 in Tab. 1). In this case, the conditions required for $R_M$ are $R_M \ll R_{out}$, to preserve the linearity, and $R_M < R_{lat}$, to ensure the recovery of the nanowire superconductivity, being $R_{lat}$ the resistance value that causes the latching of the nanowires in a stable normal state [37]. The high input resistance of the amplifier increases $\Delta R_{th}$, resulting in a total number of levels of 105 and 35 that can be equally spaced by the values $\Delta R_{2\sigma}$ and $\Delta R_{6\sigma}$, respectively.

In conclusion, we demonstrated an easy-to-use and scalable architecture for the readout of multiple superconductive detectors integrated in a high-dimensional PIC. This result has a tremendous importance in the field of photonic quantum technologies, where in experiments like Boson Sampling, Quantum Walk and Photonic Quantum Computing require the reading of a large number of modes. With the proposed approach, the position of single photons propagating in more than 100 channels can be resolved with a single coax cable. This architecture provides photon-number resolution by the engineering of the resistance values, holding an advantage over proposed readout structures. The proposed scheme allows the reduction of the thermal load inside cryostats that is a key point toward the realization of large-scale integrated experiments using SNSPDs.

AUTHOR INFORMATION

**Corresponding Author**

*alessandro.gaggero@ifn.cnr.it

**Author Contributions**

‡These authors contributed equally.

ACKNOWLEDGMENT

This work was financially supported by the projects H2020-FETPROACT-2014 no. 641039, QUCHIP, "Quantum Simulation on a Photonic Chip" and the Premiale MIUR no. 543/2015 Q-SecGroundSpace "Intermodal Secure Quantum Communication on Ground and Space". F. Martini gratefully recognizes the support of the project H2020-MSCA-IF-2017, no.795923, SHAMROCK "Superconductive MiR phOton Counter".

Reference

(1)    Knill, E.; Laflamme, R. and Milburn, G. J. A scheme for efficient quantum computation with linear optics. Nature **2001**, 409, 46.

(2)    Aspuru-Guzik and Walther, P. Photonic quantum simulators. Nat. Physics 8, 285 (2012).

(3)    Politi, A.; Cryan, M. J.; Rarity, J. G.; Yu, S. and O'Brien, J. L. Silica-on-silicon waveguide quantum circuits. Science 2008, 2, 646.

(4)    Peruzzo, A.; Lobino, M.; Matthews, J. C. F.; Matsuda, N.; Politi, A.; Poulios, K.; Zhou, X.; Lahini, Y.; Ismail, N.; Wörhoff, K.; Bromberg, Y.; Silberberg, Y.; Thompson, M. G. and O'Brien, J. L. Quantum walks of correlated photons. Science **2009**, 329, 1500–1503.

(5)    Spagnolo, N.; Vitelli, C.; Bentivegna, M.; Brod, DJ; Crespi, A.; Flamini, F.; Giacomini, S.; Milani, G.; Ramponi, R.; Mataloni, P.; Osellame, R.; Galvãõ, E. F. and Sciarrino, F. Experimental validation of photonic boson sampling. Nat. Photonics **2014**, 8, 615.

(6)    Crespi, A.; Osellame, R.; Ramponi, R.; Giovannetti, V.; Fazio, R.; Sansoni, L.; De Nicola, F.; Sciarrino, F and Mataloni, P. Anderson localization of entangled photons in an integrated quantum walk. Nat. Photonics **2013**, 7, 322–328.

(7)    Wang, J.; Paesani, S.: Ding, Y.; Santagati, R.; Skrzypczyk, P.; Salavrakos, A.; Tura, J.; Augusiak, R.; Mančinska, L.; Bacco, D.; Bonneau, D.; W. Silverstone, J.; Gong, Q.; Acín, A.; Rottwitt, K.; Oxenløwe, L. K.; O'Brien, J. L.; Laing, A. and Thompson, M. G. Multidimensional quantum entanglement with large-scale integrated optics. Science **2015**, 360, 285.

(8)    Hadfield, R. Single-photon detectors for optical quantum information applications. Nat. Photonics 2009, 3, 696.

(9)    Sprengers, J. P.; Gaggero, A.; Sahin, D.; Jahanmirinejad, S.; Frucci, G.; Mattioli, F.; Leoni, R.; Beetz, J.; Lermer, M.; Kamp, M.; Höfling, S.; Sanjines, R.; and Fiore, A.; Waveguide superconducting single-photon detectors for integrated quantum photonic circuits. Appl. Phys. Lett. 2011, 99, 181110.

(10)    Pernice, W.H.P.; Schuck, C.; Minaeva, O.; Li, M.; Goltsman, G.; Sergienko, A.V. and Tang, H.X. High-speed and high-efficiency travelling wave single-photon detectors embedded in nanophotonic circuits. Nat. Communications. **2012**, 3, 1325.

(11)    Schuck, C.; Pernice, W.H.P.; and Tang, H.X. Waveguide integrated low noise NbTiN nanowire single-photon detectors with milli-Hz dark count rate. Sci. Rep. **2013**, 3, 1893.

(12)    Sahin, D.; Gaggero, A.; Zhou, Z.; Jahanmirinejad, S.; Mattioli, F.; Leoni, R.; Beetz, J.; Lermer, M.; Kamp, M.; Höfling, S. and Fiore, A. Waveguide photon-number-resolving detectors for quantum photonic integrated circuits. App. Phys. Lett. **2013**, 103, 111116.

(13)    Najafi, F.; Mower, J.; Harris, N.C.; Bellei, F.; Dane, A.; Lee, C.; Hu, X.; Kharel, P.; Marsili, F.; Assefa, S.; Berggren, K.K. and Englund, D. On-chip detection of non-classical light by scalable integration of single-photon detectors. Nat. Communications 2013, 6, 5873.

(14)    Reithmaier, G.; Kaniber, M.;. Flassig, F.; Lichtmannecker, S.; Müller, K.; Andrejew, A.; Vuckovic, J.; Gross, R. and Finley, J. J. On-chip generation, routing, and detection of resonance fluorescence. Nano Lett. 2015, 15, 5208–5213.

(15)    Rath, P.; Kahl, O.; Ferrari, S.; Sproll, F.; Lewes-Malandrakis, G.; Brink, D.; Ilin, K.; Siegel, M.; Nebel, C. and Pernice, W. Superconducting single photon detectors integrated with diamond nanophotonic circuits. Light Sci. Appl. **2015**, 4, e338.

(16)    Vetter, A.; Ferrari, S.; Rath, P.; Alaee, R.; Kah, O.; Kovalyuk, V.; Diewald, S.; Goltsman, G. N.; Korneev, A.; Rockstuh, C. and Pernice, W. H. P. Cavity-enhanced and ultrafast superconducting single-photon detectors. Nano Lett. **2016**, 16, 7085–7092.

(17)    Allman, M. S.; Verma, V. B.; Stevens, M.; Gerrits, T.; Horansky, R. D.; Lita, A. E.; Marsili, F.; Beyer, A.; Shaw, M. D.; Kumor, D.; Mirin, R. and Nam, S. W. A near-infrared 64-pixel superconducting nanowire single photon detector array with integrated multiplexed readout. Appl. Phys. Lett. **2015**, 106, 192601.

(18)    Doerner, S.; Kuzmin, A.; Wuensch, S.; Charaev, I.; Boes, F.; Zwick, T. and Siegel, M. Frequency-multiplexed bias and readout of a 16-pixel superconducting nanowire single-photon detector array. Appl. Phys. Lett. **2017**, 111, 032603.

(19)    Hofherr, M.; Arndt, M.; Il'in, K.;, Henrich, D.; Siegel, M.; Toussaint, J.; May, T. and Meyer, H. G. Time-tagged multiplexing of serially biased superconducting nanowire single-photon detectors. IEEE Trans. Appl. Supercond. **2013**, 23, 2501205.

(20)    Zhao, Q. Y.; Zhu, D.; Calandri, N.; Dane, A. E.; McCaughan, A. N.; Bellei, F.; Wang, H. Z.; Santavicca, D. F. and Berggren, K. K. Single-photon imager based on a superconducting nanowire delay line. Nat. Photonics **2017**, 11, 247–251.

(21)    Zhu, D.; Zhao, Q. Y.; Choi, H.; Lu, T. J.; Dane, A. E.; Englund, D. and Berggren, K. K., A scalable multi-photon coincidence detector based on superconducting nanowires. Nat. Nanotechnology **2018**, 13, 596.

(22)    Divochiy, A.; Marsili, F.; Bitauld, D.; Gaggero, A.; Leoni, R.; Mattioli, F.; Korneev, A; Seleznev, V.; Kaurova, N.; Minaeva, O.; Gol'tsman, G.; Lagoudakis, K. G.; Benkhaoul, M.; Lévy, F. and Fiore, A. Superconducting nanowire photon-number-resolving detector at telecommunication wavelengths. Nature Photonics **2008**, 2, 302.

(23)    Jahanmirinejad, S.; Frucci, G.; Mattioli, F.; Sahin, D.; Gaggero, A.; Leoni, R. and Fiore, A. Photon-number resolving detector based on a series array of superconducting nanowires. Appl. Phys. Lett. **2012**, 101, 072602.

(24)    Mattioli, F.; Zhou, Z.; Gaggero, A.; Gaudio, R.; Leoni, R. and Fiore, A. "Photon-counting and analog operation of a 24-pixel photon number resolving detector based on superconducting nanowires." Opt. Express 2016, 24, 9067.

(25)    Renema, J. J.; Frucci, G.; Zhou, Z.; Mattioli, F.; Gaggero, A.; Leoni, R.; de Dood, M. J. A.; Fiore, A. and van Exter, M. P. Universal response curve for nanowire superconducting single-photon detectors. Phys Rev. B 2013, 87, 174526.

(26)    Renema, J. J.; Gaudio, R.; Wang, Q.; Zhou, Z.; Gaggero, A.; Mattioli, F.; Leoni, R.; Sahin, D.; de Dood, M. J. A.; Fiore, A. and van Exter M. P. Experimental test of theories of the detection mechanism in a nanowire superconducting single photon detector." Phys Rev. Lett. 2014, 112, 117604.

(27)    Sahin, D.; Gaggero, A., Hoang, T. B.; Frucci, G.; Mattioli, F.; Leoni, R.; Beetz, J.; Lermer, M.; Kamp, M.; Höfling, S. and Fiore, A. Integrated autocorrelator based on superconducting nanowires. Opt. Express 2013, 21, 11162.

(28)    Schwartz, M.; Schmidt, E.; Rengstl, U.; Hornung, F.; Hepp, S.; Portalupi, S. L.; llin, K.; Jetter, M.; Siegel, M. and Michler P. Fully On-Chip Single-Photon Hanbury-Brown and Twiss Experiment on a Monolithic Semiconductor−Superconductor Platform. Nano let. 2018, 18, 6892.

(29)    Cernansky, R.; Martini, F. and Politi, A. Complementary metal-oxide semiconductor compatible source of single photons at near-visible wavelengths. Opt. Letters 2018, 43 1230.

(30)    Kerman, A. j.; Dauler, E. A.; Keicher, W. E.; Yang, J. K. W.; Berggren, K. K.; Gol'tsman, G. and Voronov, B. Kinetic-inductance-limited reset time of superconducting nanowire photon counters. Appl. Phys. Let. 2006, 88, 111116.

(31)    http://www.caltechmicrowave.org/amplifiers;

(32)  $\Delta R_{th} = v_n / I_B$, with $v_n = (4k_BTBR_{out})^{1/2}$ given by the Johnson noise of the amplifier resistance at T=300K and IB=17 μA.

(33)  Ejrnaes, M.; Cristiano, R.; Quaranta, O.; Pagano, S.; Gaggero, A.; Mattioli, F.; Leoni, R.; Voronov, B. and Gol'tsman G. A cascade switching superconducting single photon detector. Appl. Phys. Lett. **2007**, 91 262509.

(34)  Marsili, F.; Najafi, F.; Dauler, E.; Bellei, F.; Hu, X.; Csete, M.; Molnar, R. J. and Berggren, K. K. Single-photon detectors based on ultranarrow superconducting nanowires. Nano Lett. **2011**, 11, 2048.

(35)  This value have been derived from the ΔRth considering at least a broadening similar to the case at 300K.

(36)  In general, $N=\sum_{n=1}^{Ch}\binom{Ch}{n}=Ch+\binom{Ch}{2}+\binom{Ch}{3}+\ldots=Ch+\frac{Ch(Ch-1)}{2}+\frac{Ch(Ch-1)(Ch-2)}{6}+\ldots=2^{Ch}-1$ is the number of levels necessary to discriminate n photons. In Tab. 1 N=Ch, as we are discussing the case n=1, i.e. the single photon regime.

(37)  Zhang, L.; You, L.; Yang, X.; Wu, J.; Lv, C.; Guo, Q.; Zhang, W.; Li, H.; Peng, W.; Wang, Z. and Xie, X. Hotspot relaxation time of NbN superconducting nanowire single-photon detectors on various substrates. Sci. Reports 2018, 8, 1486.